\documentclass{Interspeech2024}

\usepackage{lipsum}
\usepackage{multirow}
\usepackage{booktabs}

\interspeechcameraready

\title{Improving Neural Biasing for Contextual Speech Recognition by Early Context Injection and Text Perturbation}

\name[affiliation={1}]{Ruizhe}{Huang}
\name[affiliation={1}]{Mahsa}{Yarmohammadi}
\name[affiliation={1}]{Sanjeev}{Khudanpur}
\name[affiliation={2}]{Daniel}{Povey}

\address{
  $^1$Johns Hopkins University, USA\\
  $^2$Xiaomi Corp., China}
\email{\{ruizhe,mahsa,khudanpur\}@jhu.edu}


\begin{document}

\maketitle

\begin{abstract}    
    Existing research suggests that automatic speech recognition (ASR) models can benefit from additional contexts (e.g., contact lists, user specified vocabulary). Rare words and named entities can be better recognized with contexts. In this work, we propose two simple yet effective techniques to improve context-aware ASR models.
    % First, we inject contexts into the encoder at an early stage instead of merely at the last layer.
    First, we inject contexts into the encoders at an early stage instead of merely at their last layers.
    % We propose to use auxiliary CTC loss to regularize the intermediate context-injected layers.
    Second, 
    % we observe that the word error rates on the training data is already very low, even for the rare words, without providing any contexts.
    to enforce the model to leverage the contexts during training, we perturb the reference transcription with alternative spellings so that the model learns to rely on the contexts to make correct predictions. 
    % We show that our proposed techniques can significantly improve the performance of the context-aware transducer models.
    On LibriSpeech, our techniques together reduce the rare word error rate by 60\% and 25\% relatively compared to no biasing and shallow fusion, making the new state-of-the-art performance.
    On SPGISpeech and a real-world dataset ConEC, our techniques also yield good improvements over the baselines.
\end{abstract}

\section{Introduction}
\label{sec:introduction}
%
%% Transducer models have been widely used in automatic speech recognition (ASR) due to their effectiveness and streaming nature.
%% The transducer model has the ability to directly model the alignment between the input speech and the output text.

%A conventional E2E ASR model takes merely acoustics features as input and outputs the corresponding text transcription. However, human speech recognition doesn’t occur in isolation. In addition to acoustic cues, we often rely on various contextual resources, such as contact lists or user specified vocabulary, to aid in understanding and interpreting spoken content. In particular, these contextual cues play a significant role in recognizing rare words and named entities.

Human speech recognition does not occur in isolation. In addition to acoustic cues, we often rely on various contextual resources, such as semantic or visual context or speaker's background knowledge, to aid in understanding and interpreting spoken content. In particular, these contextual cues play a significant role in recognizing rare words and named entities.
End-to-end (E2E) automatic speech recognition (ASR) has emerged as the dominant solution of ASR, due to its simplicity of modeling and impressive performance. However, a conventional E2E ASR model takes merely acoustics features as input and outputs the corresponding text transcription.

Recently, various contextual biasing techniques (contextual ASR) have been proposed to improve standard E2E ASR models, 
including \cite{Fox2022ImprovingCR, Dingliwal2023PersonalizationOC, Lei2023PersonalizationOC} for connectionist temporal classification (CTC) models, 
\cite{Pundak2018DeepCE, Bruguier2019PhoebePC, Huber2021InstantOW, Zhang2022EndtoendCA} for attention-based encoder-decoder (LAS) models, \cite{Jain2020ContextualRF, Le2020DeepSF, Le2021ContextualizedSE, Chang2021ContextAwareTT, Sathyendra2022ContextualAF, Sun2022MinimisingBW, Pandey2023ProcterPC, Yang2023PromptASRFC, Futami2023PhonemeawareEF, Tang2024ImprovingAC} for transducer models and more recently \cite{Chen2023SALMSL, Sun2023ContextualBO, Lakomkin2023EndtoEndSR, Everson2024TowardsAR} for (or with) large language models (LLMs).
Following the majority of prior research, we define the context to be lists of biasing words, which are usually rare words in the model's training data. Other types of context, such as visual contexts~\cite{Pramanick2022CanVC}, date-time and location~\cite{Ray2021ImprovingRA} are out of scope of this paper.

In general, contextual ASR can be achieved either in a shallow way or deep way (or a hybrid of the two). In shallow biasing, the internal representations of the E2E models are unchanged. The contextual biasing most likely happens only during the decoding process, where the contexts are used to guide the beam search, e.g., shallow fusion~\cite{Fox2022ImprovingCR, Zhao2019ShallowFusionEC, Wang2023ContextualBW, Le2019G2GTP, Williams2018ContextualSR}, spelling correction~\cite{Wang2022TowardsCS}. Shallow fusion is considered as a simple yet robust baseline, as it can be easily integrated into beam search to provide moderate improvement in recognizing rare words. In deep (neural) biasing, the context is injected into the E2E model to edit the internal representations of the models and make a potentially new output distribution, e.g., with cross-attention over the biasing lists~\cite{Pundak2018DeepCE, Chang2021ContextAwareTT}. 
Neural biasing has been reported outperforming shallow fusion~\cite{Dingliwal2023PersonalizationOC, Pundak2018DeepCE, Le2020DeepSF, Le2021ContextualizedSE, Chang2021ContextAwareTT, Sathyendra2022ContextualAF}, due to the specialized parameters trained to accommodate contexts. However, many existing work~\cite{Zhang2022EndtoendCA, Sun2022MinimisingBW, Pandey2023ProcterPC, Yang2023PromptASRFC, Futami2023PhonemeawareEF, Tang2024ImprovingAC} do not directly compare their neural biasing approaches with shallow fusion. They only report the gains over non-contextual ASR baselines.

This paper proposes two techniques that can improve neural biasing across various ASR models.
First, we inject contexts into earlier layers of the encoder as opposed to only the last layer. Although this idea has been explored in some existing work~\cite{Wu2023DualModeNE, Yang2023PromptASRFC}, the specific layer and the number of layers to be integrated with contexts remain unclear. 
Other work~\cite{Dingliwal2023PersonalizationOC, Pandey2023ProcterPC} use the outputs from intermediate encoder layers as the queries for contexts lookup, but the resulting contextual embedding is still integrated with the encoder's final output at the last layer. While some work~\cite{Lei2023PersonalizationOC, Le2020DeepSF} raise concerns about runtime latency associated with neural biasing, we report decoding runtime in our experiments and find that the overhead of early context injection is negligible when the biasing lists are of size $500$. For larger biasing lists, light-weight algorithms may be applied to shorten the biasing lists, which can be a future work.

Secondly, during training, we propose to perturb the reference transcription with alternative, similar-sounding spellings of the rare words. For example, we opt to replace the word ``Klein'' with a random alternative spelling ``Klane'' in both the transcription and contexts. Hence, the end-to-end trained model is forced to rely on the contexts to make correct predictions. Alternative spellings has been explored in~\cite{Fox2022ImprovingCR, Lei2023PersonalizationOC, Le2020DeepSF, Le2021ContextualizedSE, Le2019G2GTP, Huang2020ClassLA, Chen2013UsingPF, Alon2018ContextualSR}. Among them,~\cite{Fox2022ImprovingCR, Lei2023PersonalizationOC, Le2019G2GTP, Huang2020ClassLA, Chen2013UsingPF} use alternative spelling during decoding to improve the recall of rare or out-of-vocabulary words. Closely related to our work are~\cite{Le2020DeepSF, Le2021ContextualizedSE, Alon2018ContextualSR} where the alternative spellings are used as data augmentation during training. However, their neural biasing architecture is very different from ours. \cite{Alon2018ContextualSR} is based on CLAS architecture~\cite{Pundak2018DeepCE} and tries to better distinguish phonetically confusable phrases. \cite{Le2020DeepSF, Le2021ContextualizedSE} use a contextual predictor (``PLM'') which is implemented by a prefix tree, instead of cross attention mechanism, in their transducer model. Thus, their encoders are not context-aware, although transducers are ``encoder-heavy'' models. Note that~\cite{Fox2022ImprovingCR, Lei2023PersonalizationOC, Le2019G2GTP, Huang2020ClassLA, Chen2013UsingPF, Alon2018ContextualSR} also propose algorithms or models to generate alternative spellings, e.g., a grapheme-to-grapheme (G2G) model~\cite{Le2019G2GTP}, which may further benefit our approach. In this work, we simply use less than $200$ hand-crafted linguistic rules (e.g., ``ein''$\leftrightarrow$``ane'', ``s''$\leftrightarrow$``z'') that cover all $26$ English letters and show this already goes a long way.

% distractors scheduler?

Despite the simplicity of our techniques, we achieve the new state-of-the-art results on LibriSpeech~\cite{Panayotov2015LibrispeechAA} in the contextual ASR setup~\cite{Le2021ContextualizedSE}. 
Furthermore, we demonstrate promising improvements over shallow fusion with neural biasing on two public datasets, SPGISpeech~\cite{ONeill2021SPGISpeech50} and ConEC~\cite{ConEC}, where the latter uses real-world contexts rather than synthesized contexts from the ground truth. Our implementation and experiment results are available in the ConEC repository\footnote{\scriptsize{\url{https://github.com/huangruizhe/ConEC}}}.
% in icefall\footnote{\scriptsize{\url{https://github.com/k2-fsa/icefall/}}} toolkit

% These results are never reported in the contextual ASR related work.
% To our best knowledge, we also report the best results on SPGISpeech and ConEC in the contextual ASR literature.

% [5, 6, 8] uses TTS to generate alternate spellings for bias terms

\vspace{-2pt}
\section{Contextual ASR}
\label{sec:contextual_asr}
In this section, we review conventional, non-contextual ASR models, taking transducers as an example. Then we describe the integration of neural contextual biasing by cross attention mechanism to the transducer models.

%\vspace{-2pt}
\subsection{Transducer ASR Model}

Transducer model is first proposed in~\cite{Graves2012SequenceTW} to learn the transformation between sequences, e.g., from speech to text transcription. Formally, it learns the probability $p(\mathbf{W}|\mathbf{X})$ of word (or word piece) sequence $\mathbf{W} = (w_1, w_2, ..., w_U)$ of length $U$, given a speech feature sequence $\mathbf{X} = (\mathbf{x}_1, \mathbf{x}_2, ..., \mathbf{x}_T)$ of length $T$. Transducer model has been widely used in ASR due to its effectiveness and streaming nature.
It has three components. 

\textbf{Encoder}: The encoder serves the role of an acoustic model. It takes the feature sequence $\mathbf{X}$ as input, and transforms it to a sequence $\mathbf{H}^{enc} = (\mathbf{h}_1^{enc}, \mathbf{h}_2^{enc}, ..., \mathbf{h}_T^{enc})$ of high-level representations of the input:
\begin{equation}
    %\vspace{-1pt}
    \mathbf{H}^{enc} = f^{enc}(\mathbf{X})
    %\vspace{-1pt}
\end{equation}
There can be many layers of the transformation. The output from the $i$-th layer $f^{enc}_i$ is denoted as $\mathbf{H}^{enc}_i$, and the final output is $\mathbf{H}^{enc}$. Note, there can be downsampling along the time axis, which is omitted here for simplicity. In practice, the encoder is implemented by a recurrent neural network (RNN) or more recently Conformer~\cite{Gulati2020ConformerCT} architecture. Encoder can take up most parameters (e.g., more than $90\%$) in the transducer model.

\textbf{Predictor}. Given $\mathbf{H}^{enc}$ from the encoder, one may produce a frame-wise softmax distribution over the output word pieces vocabulary, which is the idea of a CTC model. The transducer model, on the other hand, explicitly imposes dependencies between the output word pieces in $\mathbf{W}$ by the predictor. It acts like a language model:
\begin{equation}
    %\vspace{-1pt}
    \mathbf{h}^{pred}_{u-1} = f^{pred}(w_1, w_2, ..., w_{u-1})
    %\vspace{-1pt}
\end{equation}
\textbf{Joiner}. The joiner takes the embeddings $\mathbf{h}^{enc}_{t}$ and $\mathbf{h}^{pred}_{u}$ from both the encoder and the predictor to produce a softmax probability distribution over the output word pieces vocabulary:
\begin{equation}
    %\vspace{-1pt}
    p(w_{t, u}|\mathbf{X}) = \text{Softmax}(f^{join}(\mathbf{h}^{enc}_{t}, \mathbf{h}^{pred}_{u-1}))
    %\vspace{-1pt}
\end{equation}
Note that the output from the joiner is of the shape $(T, U, V)$ for a single input sequence, where $V$ is the size of the word pieces vocabulary. From this output, the probability of the ground-truth transcription $\mathbf{\bar{W}}$ can be computed efficiently via dynamic programming, which will be maximized during training.

\subsection{Neural Biasing with Cross Attention}
\label{sec:neural-biasing}

CLAS~\cite{Pundak2018DeepCE} was first proposed to bias the attention-based encoder-decoder (LAS) models with cross attention. Later,~\cite{Jain2020ContextualRF, Chang2021ContextAwareTT, Sathyendra2022ContextualAF} applied cross attention to bias transducer models, which has become a popular solution recently. In general, contextual ASR models learn to predict the probability $p(\mathbf{W}|\mathbf{X}, \mathbf{C})$ of word sequences given both acoustic feature sequence $\mathbf{X}$ and some context $\mathbf{C}$. In this work, we consider context for each utterance as a list of biasing words or phrases $\mathbf{C} = \{\mathbf{c}_0, \mathbf{c}_1, \mathbf{c}_2, ..., \mathbf{c}_N\}$. Note, $\mathbf{c}_0$ is a special entry corresponding to the {\em no-bias} option.

Intuitively, when we predict the word piece at the $t$-th frame, we expect the relevant entries in the biasing list get sufficient attention. To do this, one can first independently represent each word or phrase (of various number of characters) $\mathbf{c}_j \in \mathbf{C}$ by a fixed-dimensional embedding $\mathbf{c}_j^e \in \mathbf{C}^e$ of $D$ dimensions. An LSTM \textbf{context encoder} can be employed to accomplish this:
\begin{equation}
    \mathbf{c}_j^{e} = \text{BiLSTM}(\mathbf{c}_j)
\end{equation}
Note $\mathbf{c}_0^{e}$ is hard-wired to all zeros.

Then, for each frame $\mathbf{h}_{t}^{enc}$ from the encoder output --- it can also be the output from the $i$-th encoder layer, we omit the subscript $i$ for simplicity --- or for each frame $\mathbf{h}^{pred}_{u-1}$ from the predictor's embedding, $\mathbf{h}_{t}^{enc}$ or $\mathbf{h}^{pred}_{u-1}$ is used as the query to attend to the embedded biasing list, whose keys and values are both $\mathbf{c}_j^e \in \mathbf{C}^e$. This makes the cross-attention \textbf{contextual biasing adapter}:
\begin{equation}
    \mathbf{A}^{enc} = \text{MHA}(\text{q=}\mathbf{H}^{enc}, \text{k=}\mathbf{C}^e, \text{v=}\mathbf{C}^e)
\end{equation}
where $\mathbf{A}^{enc}$, $\mathbf{H}^{enc}$, $\mathbf{C}^e$ are matrices of shape $T \times N$, $T \times D$ and $N \times D$ respectively (additional projection layers are omitted here). For the $t$-th frame, $\mathbf{a}_{t}^{enc} \in \mathbf{A}_{t}^{enc}$ defines an attention weights distribution over the $N$ biasing words. Their weighted sum is the attention output: $\mathbf{b}_{t}^{enc} = {\mathbf{a}_{t}^{enc}}^\top \cdot \mathbf{C}^e$.

\begin{figure}[t]
  \centering
  \includegraphics[width=0.9\linewidth, trim=1.2cm 0.6cm 0cm 0cm, clip]{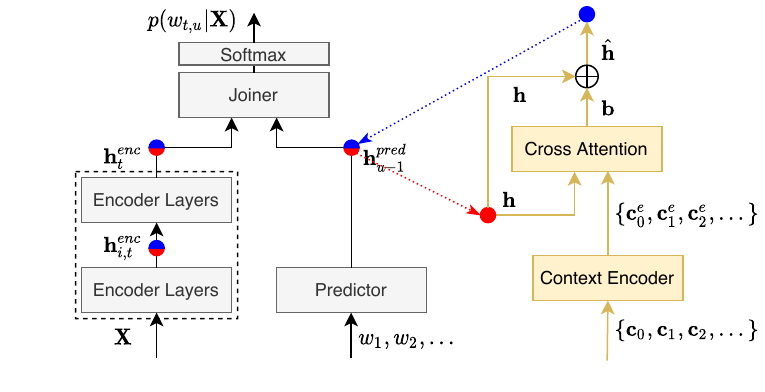}
  %\vspace{-2pt}
  \caption{The transducer model (left) and its contextual biasing module (right). The red/blue dots mark the locations where the contexts can be injected to the main model, by plugging in the contextual biasing module. During training, the gray modules can be frozen, while only the yellow modules are trained.}
  % to the model's internal representations
  \label{fig:transducer}
  \vspace{-12pt}
\end{figure}

Finally, the attention output $\mathbf{b}_{t}^{enc}$ is used to ``edit'' the encoder's (or decoder's) embedding. This can be implemented via element-wise addition:
\begin{equation}
    \hat{\mathbf{h}}_t^{enc} = \mathbf{h}_t^{enc} + \mathbf{b}_{t}^{enc}
\end{equation}
Thus, the internal representations of the transducer model become context-aware. Modified embeddings $\hat{\mathbf{h}}_t^{enc}$ and $\hat{\mathbf{h}}_{u-1}^{pred}$ are fed to the next encoder layers or the joiner to make context-aware predictions. This is illustrated in Figure~\ref{fig:transducer}.

The above context encoder and the cross-attention biasing adapter can be trained in an end-to-end fashion together with the transducer model. In~\cite{Jain2020ContextualRF, Chang2021ContextAwareTT}, all parameters are trained from scratch. In~\cite{Sathyendra2022ContextualAF}, the transducer's parameters are pretrained and frozen. Only a small number of the rest of the parameters (yellow modules in Fig.~\ref{fig:transducer}) are updated. The benefit is that the neural biasing maintains the performance of the pretrained model when no context is provided. We follow~\cite{Sathyendra2022ContextualAF} in this paper.
% only the parameters of the context encoder and the cross-attention biasing adapter are updated.

\section{Proposed Approaches}
\label{sec:proposed}
\subsection{Early Context Injection}

In~\cite{Dingliwal2023PersonalizationOC, Pundak2018DeepCE, Bruguier2019PhoebePC, Jain2020ContextualRF, Chang2021ContextAwareTT, Sathyendra2022ContextualAF, Pandey2023ProcterPC, Tang2024ImprovingAC}, cross-attention neural biasing architectures are used as described in section~\ref{sec:neural-biasing}. 
All of them inject contexts only into the encoder's {\em final} output. The drawback is that the contextualization of a frame $\hat{\mathbf{h}}_t^{enc}$ has no impact on the other frames, as they are edited essentially independently from one another. On the other hand, if contexts are injected to earlier encoder layers (e.g., at $\mathbf{h}_{i,t}^{enc}$ in Figure~\ref{fig:transducer}), it may have far-reaching impacts on the model's internal states via its self-attention mechanism.

One might argue that cross-attention context lookup can introduce significant computation overhead, hence they perform it only once at the last encoder layer. However, as we will show later in Section~\ref{sec:exp}, this overhead is negligible. In fact, context injection can be viewed as simply adding an extra layer to the encoder. For ASR, the input or intermediate sequences usually have several hundred frames. Thus, each layer of the encoder, e.g., a conformer~\cite{Gulati2020ConformerCT}, needs to do self-attention over several hundred items. This is a comparable computation to contexts lookup if we have several hundred biasing words to attend to.

The predictor of the transducer is usually implemented by a simple LSTM network or even a stateless $n$-gram feed-forward network~\cite{Ghodsi2020RnnTransducerWS}. Thus, we only bias the predictor's final output.

\subsection{Text Perturbation with Alternative Spellings}

When transcribing unfamiliar names of people, locations or products, humans tend to choose the most familiar or phonetically similar option. Given a reference guide, they will likely refer to it for guidance. Moreover, when the contents in the reference guide change, they adapt accordingly. Text perturbation follows this idea to train the model to optimally use the contextual information. While acoustic data augmentation (e.g.,~\cite{Park2019SpecAugmentAS}) is a common practice for ASR, text perturbation has not been widely adopted yet.

\begin{table}[b]
\centering
\vspace{-4pt}
\caption{Librispeech / SPGISpeech training data distributions}
%\vspace{-4pt}
\label{tab:traing-data}
\scriptsize
\begin{tabular}{lcc}
\hline
\rule{0pt}{2.4ex} & \textbf{LibriSpeech} & \textbf{SPGISpeech} \\ \hline
\rule{0pt}{2.4ex}Duration & 1000 hours & 5000 hours \\
\rule{0pt}{2.4ex}\# Words & 28,210,665 & 141,450,888 \\
\rule{0pt}{2.4ex}\% Rare words (RW) & 10.09\% & 6.12\% \\
\begin{tabular}[c]{@{}l@{}}Training WER \\ (common/rare)\end{tabular} & \begin{tabular}[c]{@{}c@{}}\rule{0pt}{2.4ex}5.77 \% \\ 
(5.12\%/11.57\%)\end{tabular} & \begin{tabular}[c]{@{}c@{}}\rule{0pt}{2.4ex}1.39\% \\ 
(1.27\%/3.09\%)\end{tabular} \\ 
\hline
\rule{0pt}{2.4ex}Avg utterance length & 33.5 words & 24.0 words \\
% \begin{tabular}[c]{@{}l@{}}\% utterances - \\ with rare words\end{tabular} & 91.24\% & 70.26\% \\
\rule{0pt}{2.4ex}\% Utterances: & & \\
\rule{0pt}{2.4ex}$\bullet$ with RW & 91.24\% & 70.26\% \\
\rule{0pt}{2.4ex}$\bullet$ with mis-recog RW & 29.38\% & 4.42\% \\
\begin{tabular}[c]{@{}l@{}}\rule{0pt}{2.4ex}$\bullet$ with mis-recog RW \\ $\,\,\,\,\,$(after perturbation)\end{tabular} & 90.13\% & 36.37\% \\
\hline
\end{tabular}
\end{table}

On the other hand, we observe that ASR models can overfit the training data. The word error rates on the training data is so low that the end-to-end training of neural biasing modules may not have enough chances to learn to attend to the contexts. On LibriSpeech and SPGISpeech (Section~\ref{sec:exp-data}), the training data distribution and word error rates are listed in Table~\ref{tab:traing-data}. Even without contexts, there is only 29.38\% and 4.42\% utterances in the training data containing at least one mis-recognized rare word (defined in Section~\ref{sec:exp-data}). This implies that {\em only} these utterances may potentially benefit from the contexts containing ground-truth rare words during training. After random text perturbation (details can be found in Section~\ref{sec:exp-data}), the percentage of such mis-recognized utterances containing rare words increases to 90.13\% and 36.37\% of all utterances. 
% Due to the randomness, there is more coverage of different rare words across training epochs.
% In fact, as the perturbation is random, every utterance has a chance to b

% only 1\% and 0.19\% words are mis-recognized rare words (e.g., $10.09\% \times 11.57\%$). 
% In our experiments, we increase these ratios to 2\% and 1.7\% by random perturbation of rare word spellings (with probability 20\% and 30\%). Although they are still small, they have a much bigger coverage of different rare words due to the randomness. Moreover, 

In our experiments, we apply hand-crafted linguistic rules\footnote{\tiny{\url{https://gist.github.com/huangruizhe/dd75cf44bde12751500b8c43c73f3f22}}} to obtain similar-sounding spelling alternatives. 
We replace the ``maximal'' matched pattern with its counterpart (e.g., ``lee'' $\rightarrow$ ``li'' although pattern ``e''$\leftrightarrow$``a'' is also a match).
For some languages, e.g., Chinese, it is even easier to obtain spelling alternatives by pronunciation dictionary lookup for characters.

\section{Experiments}
\label{sec:exp}
% TODO: what's rare words? Why do we put them in the context?

\subsection{Datasets}
\label{sec:exp-data}

We use LibriSpeech~\cite{Panayotov2015LibrispeechAA} and SPGISpeech~\cite{ONeill2021SPGISpeech50} to train contextual ASR models. The orthographically normalized version of SPGISpeech is used. Dataset statistics can be found in Table~\ref{tab:traing-data}. We follow the same setup as in~\cite{Le2021ContextualizedSE, Tang2024ImprovingAC} to generate artificial biasing lists during training. More specifically, we define rare words to be the words beyond top 5k/3k most frequent words for LibriSpeech/SPGISpeech. Due to the long tail nature, these words has poor ASR performance compared to the common words (Table~\ref{tab:exp-librispeech} row 1). Thus, we provide the ASR model with such words, as well as some distractors, as the context, aiming to enhance the model's ability to recognize these words. For each utterance, we include all rare words from its reference transcription into its context. We also add 100 distractors sampled from all the rare words in the training vocabulary. We observe no gains when adding more distractors.

For evaluation, we use the test sets in LibriSpeech and SPGISpeech. The LibriSpeech contexts are predefined in~\cite{Le2021ContextualizedSE}, and we define the SPGISpeech contexts similary. We also use ConEC~\cite{ConEC}, which consists of earnings calls in Earnings-21~\cite{Rio2021Earnings21AP} and their real-world supplementary materials including presentation slides, earnings news release, a list of meeting participants' names and affiliations. We follow the evaluation metrics in~\cite{Le2021ContextualizedSE, Tang2024ImprovingAC} to compute overall word error rate (WER) and the WER for common words and rare words (U-WER, B-WER). For ConEC, we also report the WERs for named entities.

For text perturbation, we randomly perturb the spellings of rare words with a given probability (0.2 for both datasets). For SPGISpeech, we also randomly (with probability of 0.8) discard utterances containing no rare words during training. 
% This makes

% In our experiments, we increase these ratios to 2\% and 1.7\% by random perturbation of rare word spellings (with probability 20\% and 30\%). Although they are still small, they have a much bigger coverage of different rare words due to the randomness. Moreover, 

\subsection{Model}

We use stateless transducer~\cite{Ghodsi2020RnnTransducerWS} with Zipformer~\cite{Yao2023ZipformerAF} encoder. The model has 15 encoder layers and 71.5M parameters in total. The transducers are trained on LibriSpeech or SPGISpeech. Then, we freeze their parameters. We use a BiLSTM context encoder of two layers and 128-dim hidden states, which is shared across all contextual biasing modules. We do not notice improvements using separate context encoders or using BERT~\cite{Devlin2019BERTPO} as the encoder. Each biasing adapter is implemented by a 4-head, 128-dim multi-head dot product attention layer, as well as necessary projection layers. Overall, the contextual biasing modules account for 3.7\%--6.7\% of the parameters compared to the transducer model. Our main contextual ASR model injects contexts at both the 9th and 15th (the last) encoder layers.
The model is optimized with ScaledAdam~\cite{Yao2023ZipformerAF} for 30 epochs.

% The context encoders are shared across all contextual biasing modules. We on't see improvement using seperate encoders or using BERT as the encoder. The encoder has most likely learned to encode the pronunciation.

\subsection{Results}

\begin{table*}[t]
  \centering
  \caption{\small Contextual ASR on LibriSpeech. Each cell is formatted as WER (U-WER / B-WER). $N$ is the number of distractors added to the biasing words list. NO (no biasing), SF (shallow fusion), NB (neural biasing), TP (text perturbation), ``@Layers: 9,15'' means to which layers contexts are injected. $\dagger$: This row shares the same results as there is no context for ``no biasing'', so $N$ is irrelevant.}
  \label{tab:exp-librispeech}
  \makebox[\textwidth][c]{
  \begin{tabular}{lccccccc} % Adjust the number of 'c' based on the desired number of columns
    \toprule
    & \multicolumn{2}{c}{N=100} & \multicolumn{2}{c}{N=500} & \multicolumn{2}{c}{N=1000} \\ % Remove N=2000 columns
    \cmidrule(lr){2-3} \cmidrule(lr){4-5} \cmidrule(lr){6-7} % Remove N=2000 columns
    & test-clean & test-other & test-clean & test-other & test-clean & test-other \\
    \midrule
    NO & \scriptsize $\dagger$ & \scriptsize $\dagger$ & \scriptsize 2.17 (1.25/9.65) & \scriptsize 5.22 (3.32/21.83) & \scriptsize $\dagger$ & \scriptsize $\dagger$ \\
    SF & \scriptsize 1.49 (1.18/3.98) & \scriptsize 4.01 (3.27/10.50) & \scriptsize 1.59 (1.26/4.22) & \scriptsize 4.11 (3.32/11.05) & \scriptsize \textbf{1.63} (1.31/\textbf{4.27}) & \scriptsize 4.26 (3.47/11.21) \\ \hline
    PromptASR~\cite{Yang2023PromptASRFC} & \scriptsize 1.73 ( - / - ) & \scriptsize 4.07 ( - / - ) & \scriptsize 2.0 ( - / - ) & \scriptsize 4.45 ( - / - ) & \scriptsize 2.13 ( - / - ) & \scriptsize 4.67 ( - / - ) \\
    Guided attn~\cite{Tang2024ImprovingAC} & \scriptsize 2.2 (1.8/5.1) & \scriptsize 5.4 (4.7/12.2) & \scriptsize n/a & \scriptsize n/a & \scriptsize 2.4 (1.9/6.4) & \scriptsize 6.0 (5.0/15.3) \\ \hline
    NB & \scriptsize 1.72 (1.17/6.03) & \scriptsize 4.13 (3.18/12.47) & \scriptsize 1.72 (\textbf{1.19}/6.06) & \scriptsize 4.33 (3.24/13.85) & \scriptsize 1.78 (\textbf{1.19}/6.60) & \scriptsize 4.34 (\textbf{3.19}/14.41) \\
    $\,\,\,$+ @Layers: 9,15 & \scriptsize 1.52 (1.13/4.65) & \scriptsize 3.69 (\textbf{3.06}/9.16) & \scriptsize 1.71 (1.21/5.75) & \scriptsize 4.00 (\textbf{3.16}/11.44) & \scriptsize 1.86 (1.28/6.58) & \scriptsize 4.38 (3.35/13.46) \\
    $\,\,\,\,\,\,$+ TP & \scriptsize \textbf{1.24} (\textbf{1.11}/\textbf{2.29}) & \scriptsize \textbf{3.32} (3.08/\textbf{5.46}) & \scriptsize \textbf{1.50} (1.24/\textbf{3.56}) & \scriptsize \textbf{3.69} (3.18/\textbf{8.19}) & \scriptsize 1.74 (1.33/5.10) & \scriptsize \textbf{4.24} (3.49/\textbf{10.84}) \\
    \bottomrule
  \end{tabular}
  }
\end{table*}

The WER results for LibriSpeech are reported in Table~\ref{tab:exp-librispeech}. When there is no context or biasing, the rare word WER can be as high as 21.83\%. Shallow fusion with biasing words provides significant improvement on this dataset, which is nearly 50\% relative WER reduction. Note that shallow fusion is insensitive to the size $N$ of the biasing words lists. The neural biasing results from two recently published papers~\cite{Yang2023PromptASRFC, Tang2024ImprovingAC} and our vanilla neural biasing implementation do not outperform our shallow fusion baseline, even though their ``no biasing'' results (not shown here) are very close to ours. When contexts are injected to both the 9th and 15th layers, we see improvements over the vanilla neural biasing. When we further perturb the reference transcriptions, we achieve the best neural biasing results, with 60\% and 25\% B-WER reduction over no biasing and shallow fusion, without degrading U-WER. Also note that neural biasing is more sensitive to the biasing list size $N$ compared to shallow fusion.

On SPGISpeech (Table~\ref{tab:spgi-conec}), with a test set of 100 hours long, our techniques again outperform the baselines. On ConEC, our methods surpass the baseline except for one entity class.

\begin{table}[h]
  \centering
  \caption{Contextual ASR on SPGISpeech and ConEC}
  \label{tab:spgi-conec}
  \scriptsize
  \begin{tabular}{lcccc}
    \toprule
    \textbf{SPGI} & \multicolumn{2}{c}{\textbf{N=100}} & \multicolumn{2}{c}{\textbf{N=500}} \\ 
    \midrule
    NO & \multicolumn{4}{c}{\scriptsize 2.10 (1.81/6.60)} \\
    SF & \multicolumn{2}{c}{\scriptsize 1.85 (1.73/3.57)} & \multicolumn{2}{c}{\scriptsize 1.85 (1.74/3.61)} \\
    NB & \multicolumn{2}{c}{\scriptsize 1.78 (1.68/3.24)} & \multicolumn{2}{c}{\scriptsize 1.82 (1.69/3.76)} \\
    +9,15,TP & \multicolumn{2}{c}{\textbf{\scriptsize 1.71 (1.65/2.56)}} & \multicolumn{2}{c}{\textbf{\scriptsize 1.79 (1.68/3.35)}} \\
    \midrule
    \textbf{ConEC} & \textbf{WER} & \textbf{PERSON} & \textbf{PRODUCT} & \textbf{ORG} \\
    \midrule
    NO & \scriptsize 10.41 (8.71/26.02) & \scriptsize 45.9 & \scriptsize 24.25 & \scriptsize 29.54 \\
    SF & \scriptsize \textbf{10.29} (\textbf{8.70}/24.84) & \scriptsize 39.82 & \scriptsize \textbf{21.86} & \scriptsize 26.09 \\
    NB & \scriptsize 10.66 (8.93/26.44) & \scriptsize 41.38 & \scriptsize 24.85 & \scriptsize 27.46 \\
    +9,15,TP & \scriptsize 10.40 (8.76/\textbf{24.61}) & \scriptsize \textbf{35.72} & \scriptsize 25.48 & \scriptsize \textbf{25.70} \\
    \bottomrule
  \end{tabular}
\end{table}

\begin{table}[h]
\centering
\caption{Context injection to different encoder layers on LibriSpeech ($N=500$)}
\label{tab:layers}
\scriptsize
\begin{tabular}{lcc}
\toprule
\textbf{@Layers} & \textbf{WER (test-other)} & \textbf{Runtime (min)} \\
\midrule
NO & \scriptsize 5.22 (3.32/21.83) & \scriptsize 2.6 \\
15 (the last) & \scriptsize 4.33 (3.24/13.85) & \scriptsize 3.2 \\
9 & \scriptsize 4.25 (3.25/13.03) & \scriptsize 3.3 \\
6, 15 & \scriptsize 4.19 (3.24/12.55) & \scriptsize 3.5 \\
9, 15 & \scriptsize \textbf{4.00 (3.16/11.44)} & \scriptsize 3.4 \\
11, 15 & \scriptsize 4.14 (3.24/12.09) & \scriptsize 3.3 \\
6, 9, 11, 15 & \scriptsize 4.19 (3.23/12.63) & \scriptsize 3.6 \\
\bottomrule
\end{tabular}
\end{table}

Next, we explore which layers are optimal for integrating the contexts (Table~\ref{tab:layers}). Text perturbation is disabled here. It appears that the combination of layers 9 and 15 yields the best performance. We also measure the wall-clock time to decode LibriSpeech {\em test-other} (of 5.3 hours) with a beam size of 4 for beam search. It takes about 3.5 minutes on one NVIDIA Tesla V100 GPU, even when we inject contexts to 4 encoder layers. 
As a reference, it takes 2.6 minutes for a non-contextual transducer model to decode the same data.

Finally, we search for the best probability for perturbing each rare word (Table~\ref{tab:prob}). When the probability takes 0.2, the contextual ASR model has a balanced WER performance for common and rare words.

\begin{table}[h]
\centering
\caption{The impact of the probability for text perturbation on LibriSpeech ($N=500$, @Layers: 9,15)}
\label{tab:prob}
\scriptsize
\begin{tabular}{lcc}
\toprule
& \textbf{test-clean} & \textbf{test-other} \\
\midrule
NO  & 2.17 (1.25/9.65) & 5.22 (3.32/21.83) \\
0.1 & 1.60 (1.27/4.25) & 3.89 (3.38/8.34) \\
0.2 & \textbf{1.50 (1.24/3.56)} & \textbf{3.69} (\textbf{3.18}/8.19) \\
0.4 & 1.65 (1.40/3.66) & 3.81 (3.45/\textbf{6.99}) \\
\bottomrule
\end{tabular}
\end{table}

% Table 4: layers, runtime
% Table 5: probabilities

\section{Conclusion}
\label{sec:conclusions}
In this paper, we apply two techniques to improve cross attention based neural biasing for contextual ASR. First, we inject contexts into intermediate encoder layers in addition to the last layer. Second, during training, we replace the rare words with their similar-sounding alternative spellings in both the reference transcription and contexts. The techniques yield significant improvement in recognizing rare words and named entities on three datasets, including a real-world contextual ASR test set. Future work may explore using advanced alternative spellings generators, shortening the biasing lists or reducing the sensitivity of contextual ASR models to the distractors.

\section{Acknowledgements}
The authors would like to acknowledge the helpful discussion and support from Jing Liu, Mingzhi Yu, Grant Strimel, and Ariya Rastrow. This work was supported by a fellowship from JHU + Amazon Initiative for Interactive AI (AI2AI).

\bibliographystyle{IEEEtran}
\bibliography{mybib}

\end{document}